# Cooperative engineering the multiple radio-frequency fields to reduce the X-junction barrier for ion trap chips

Yarui Liu[1,†], Zhao Wang[2,3,4,†,*] , Zixuan Xiang[1], Qikun Wang[1], Tianyang Hu[1] & Xu Wang[1,*]

[1]College of Big Data and Information Engineering, Guizhou University, Guiyang 550025, China [2]Shenzhen Institute for Quantum Science and Engineering, Southern University of Science and Technology, Shenzhen 518055, China [3]International Quantum Academy, Shenzhen 518048, China [4]Guangdong Provincial Key Laboratory of Quantum Science and Engineering, Southern University of Science and Technology, Shenzhen 518055, China

[†]These authors contributed equally to this work.

[*]E-mails: joeshardow@gmail.com (Zhao Wang), xuwang@gzu.edu.cn (Xu Wang)

**With the increasing number of ion qubits and improving performance of sophisticated quantum algorithms, more and more scalable complex ion trap electrodes have been developed and integrated. Nonlinear ion shuttling operations at the junction are more frequently used, such as in the areas of separation, merging, and exchanging. Several studies have been conducted to optimize the geometries of the radio-frequency (RF) electrodes to generate ideal trapping electric fields with a lower junction barrier and an even ion height of the RF saddle points. However, this iteration is time-consuming and commonly accompanied by complicated and sharp electrode geometry. Therefore, high-accuracy fabrication process and high electric breakdown voltage are essential. In the current work, an effective method was proposed to reduce the junction's pseudo-potential barrier and ion height variation by setting several individual RF electrodes and adjusting each RF voltage amplitude without changing the geometry of the electrode structure. The simulation results show that this method shows the same effect on engineering the trapping potential and reducing the potential barrier, but requires fewer parameters and optimization time. By combining this method with the geometrical shape-optimizing, the pseudo-potential barrier and the ion height variation near the junction can be further reduced. In addition, the geometry of the electrodes can be simplified to relax the fabrication precision and keep the ability to engineer the trapping electric field in real-time even after the fabrication of the electrodes, which provides a potential all-electric degree of freedom for the design and control of the two-dimensional ion crystals and investigation of their phase transition.**

# INTRODUCTION

Ion traps are very promising physical systems in quantum computing that could scale the number of qubits with high fidelity[1,2]. The charged atomic internal state is used to store quantum information, and the motion state is adopted for realizing multiqubit quantum gate operation[3,4]. Ion qubits are endowed with excellent coherence properties[5,6] and connectivity[7], and can be initialized and measured with near 100% efficiency[8,9], rendering them to achieve high-fidelity quantum gate operations[10–12] and demonstrate quantum error correction[13,14].

Since the ability of quantum computing depends on the number of qubits that can be manipulated with high fidelity[15,16], the methods of increasing the high-fidelity qubit number have become the key problem of the current ion trap quantum computing researches[17,18]. In the study of ion traps, a chain of linear ions is widely employed to investigate the properties of quantum computing. As the number of ions increases, the linear ion chain will be faced with a nonuniform distribution, which will lead to the problem that it is of great difficulty to address individually on each ion[19]. Besides, there are a series of issues arising simultaneously, such as the reducing stability of ion crystals[20], the increasing density of motional modes, enhancing crosstalk between ions[1] as well as reducing fidelity of quantum logic operations[21,22].

To further increase the number of ion qubits that can be manipulated at the same time, in 2002, D. Wineland proposed a method to realize the two-dimensional expansion of trapped ions with the adoption of a quantum charge-coupled device (QCCD) structure[23–25]. By separating the zones of different functions, a few ions can be manipulated with high fidelity in a single zone[26], while multi-zones operating simultaneously increase the number of the overall qubits, preventing the fast growth of decoherence[27–31], and maintaining the high fidelity of the quantum state operation[23,32,33]. The QCCD-structured ion chip method could potentially solve the scale problem. By employing the micro-fabricating technology, various complex electrode patterns can be designed and implemented on a small semiconductor chip. Ions are shuttled by a series of time-varying voltage waveforms, which are applied to direct-current (DC) electrodes and transport the quantum information between different zones on two-dimensional QCCD-structured ion traps[2,34–36]. Consequently, the "junction" is of great importance for realizing the separating, merging and swapping operation with a lower ion loss rate, lower phonon excitation rate and fast transport speed[24,37]. However, the design of the electrodes with a junction is different from that of the linear trap; it breaks the translational symmetry, so a pseudo-potential barrier and the height of ions are both affected by the interferences from other shuttling path electrodes.

In the previous junction design methods, most of the researchers chose to optimize the radio-frequency (RF) electrode geometry and applied a single RF-driven signal[25,26,37–43]. Up till now, few studies have proposed the ion shuttling scheme by setting up several RF electrodes and controlling multiple independent RF voltages on each electrode in real-time, therefore the ions are always in the trapping saddle point[44,45]. This method requires a complex and real-time RF voltage waveform to match the ion spatial position exactly during the shuttling process, which is not easy to be realized experimentally[37,40]. However, this multi-RF voltage method is a potential way to shuttle ions through junctions without any micro-motion or phonon excitation[44–47]. To the best of our knowledge, such a method has not yet been successfully applied to the ion trap junction design.

In this study, a static multi-RF field loading protocol was put forward by optimizing the RF voltage amplitude on each RF electrode and the segmented junction RF electrode design for optimizing the junction pseudo-potential. This protocol can simplify the controlling process by separating the RF trapping potential shaping and the DC shuttling waveform design process, with the ability to optimize the junction pseudo-potential specifically according to the ion shuttling direction along the linear forward path or the corner turning. Furthermore, when combined with the electrode geometrical optimization method, the pseudo-potential barrier can be further reduced and the ion height variation tends to be less and even.

## THE MULTIPLE RADIO FREQUENCY FIELDS OPTIMIZATION

One simple and intuitive way to describe the effects of the average force of the alternating current (AC) electric field is pseudo-potential[48]. The ideal RF trapping pseudo-potential along the junction and the shuttling path is a pseudo-potential tube in which all the points are of zero value. The height of the saddle points and the ion motional frequency are even, especially in the junction central region. However, the crossed structure of the RF electrodes in the ion chip junction will lead to the generation of pseudo-potential barriers and gradients[26,39,41], which are very harmful to fast and low heating rate shuttling. The optimization of the pseudo-barrier will reduce the pseudo-potential gradient accordingly, so minimizing the potential barrier is the most important goal of this study. Other physical considerations can be found in the Supplementary Materials or in refs.[25,41].

Applying an RF voltage with the same frequency, phase, and amplitude to electrodes with a defined geometry can generate one or more saddle points in space with zero pseudo-potential, which are connected into different shapes. The ions are subjected to the RF field and the DC field interaction, forming the stable ion crystal. However, using multiple RF voltages with the same frequency, phase, and different amplitudes can shift the saddle points and engineer the pseudo-potential tube shapes without changing the geometry of the electrodes or requiring specific shapes of the electrodes[46,47], which not only caters for the linear trap but also for the trap with junctions, to reduce the RF barrier and saddle point variation. This will relax both the design complexity and the fabricating accuracy of the electrode pattern, or even reduce the chance of electrode discharge or breakdown.

The process of the multi-RF field optimization method is: A) dividing the single RF electrode into several segments. B) Setting the RF voltage on each RF segment and calculating the objective function value, like the barrier of the pseudo-potential, etc. C) Changing the RF voltage on each RF segment to minimize the objective function. In most situations, the objective function value will reduce largely to a specific value. The corresponding voltages are the optimal voltage distribution for this special division of the RF electrode. D) If the objective function has not reached expectations, the method of RF electrode segmentation can be further changed (step A) and the above voltage optimization steps can be further repeated (steps B and C).

Since the voltage distribution can still be changed even after the electrode pattern is fixed, it is possible to optimize the pseudo-potential for special ion shuttling, like corner turning or linear transport, as discussed in the next section. Furthermore, this ability to engineer the electric field in real time may help study the two-dimensional ion crystal phase transition[49].

Without loss of generality, taking the standard X-junction as an optimization study case, the segmentation of X-junction electrodes is shown in Fig. 1. The X-junction exhibits 4 shuttling paths, where each path is labeled by a letter $x$, $x \in \{A, B, C, D\}$ or $\{a, b, c, d\}$. The original single RF electrode is divided into several sub-RF electrodes (segments) and labeled as $RFix$, where the number $i$ is the sequence number of the segmental electrode in each path, $i \in \{1, 2, 3\}$; $x$ stands for the path where the sub-electrode is located. The RF voltage will be implemented at the same amplitude if the segment exhibits the same label. The center four square segments are labeled as $\{e\}$ since the segment shape is different and they are the nearest electrodes from the junction center. The coordinate system is built from the junction center (origin point), and the horizon $x$-axis and the vertical $y$-axis are drawn in the middle black color; the z direction is perpendicular to the outside of the study.

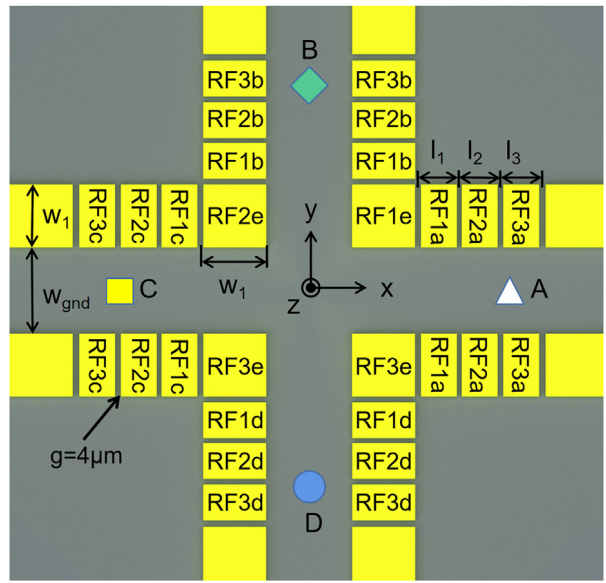

**Fig. 1 | The segmented RF electrodes are near the X-junction.** The original junction structure is based on the five-wire structure, where the RF electrode is yellow in color and the ground electrode is grey. To emphasize the RF electrodes and voltage effects, the DC electrodes are omitted. The width of the RF electrode is $w_1$. The length of each segment is $l_i$, where, the i stands for the segmental sequential number. The width of the center ground electrode is $w_{gnd}$, and all the gap of the electrodes is g. For the numerical simulation in the next paragraph, the RF electrode dimensions are chosen as $w_1 = 80$ μm, $l_i = 45$ μm, $w_{gnd} = 100$ μm, $g = 4$ μm. RF, radio frequency; DC, direct-current.

The $^{171}$Yb$^+$ ion and 30 MHz RF drive frequency are adopted as the simulating parameters to verify the effectiveness of the multi-RF optimization method, simulate and compare the trapping potential in different RF voltage distributions. Before the RF voltage optimization, all the RF electrode segments voltages are set to 100 V. Since the shape of the electrode is symmetrical about the planes $x = 0$ and $y = 0$, and the shuttling ions are always in the symmetrical plane, the simulation can be performed in zone A and in the ZOX ($y = 0$) plane solely, in a range of $x \in [0, 500]$ μm from the junction center. The unoptimized trapping pseudo-potential is simulated as a comparable reference, and the result is shown as the contour map in Fig. 2a. The pseudo-potential tube represented by dark purple blue is interrupted in the $x = 100$ μm region, forming the pseudo-potential barrier in Fig. 2d light blue line. The maximum pseudo-potential is about 5.265 meV. The maximum ion height variation of the pseudo-potential tube center (the RF saddle points) is about 58.44 μm. As the $x$ coordinate approaches zero, the ion height goes higher (solid blue line in Fig. 2e).

The method of the multi-RF field will optimize and suppress the pseudo-potential barrier largely, either for the corner shuttling case or for the linear shuttling case. For the case of corner turning shuttling, without loss of generality, it is assumed that ions move from zone A to zone B. The pseudo-potential barrier is optimized by multi-RF field optimization, and the result is shown in Fig. 2b. The corresponding RF voltage distribution is shown in Fig. 3a. The resulting RF voltages and the trapping potential are symmetrical about $x = y$. The center of the pseudo-potential tube is in the shape of a curve, and the saddle points circumvent the center of the junction. The position of the saddle point on the symmetrical $x = y$ plane is near $(x, y) = (20, 20)$ μm. Thus, in the range of $x \in [0, 20]$ μm, there is no saddle point data in the ZOX ($y = 0$) plane. Compared to the unoptimized results, the maximum pseudo-potential value has been greatly suppressed to 1.164 meV (grey solid line in Fig. 2d), and the maximum ion height variation has also been decreased to 54.19 μm (grey solid line in Fig. 2e).

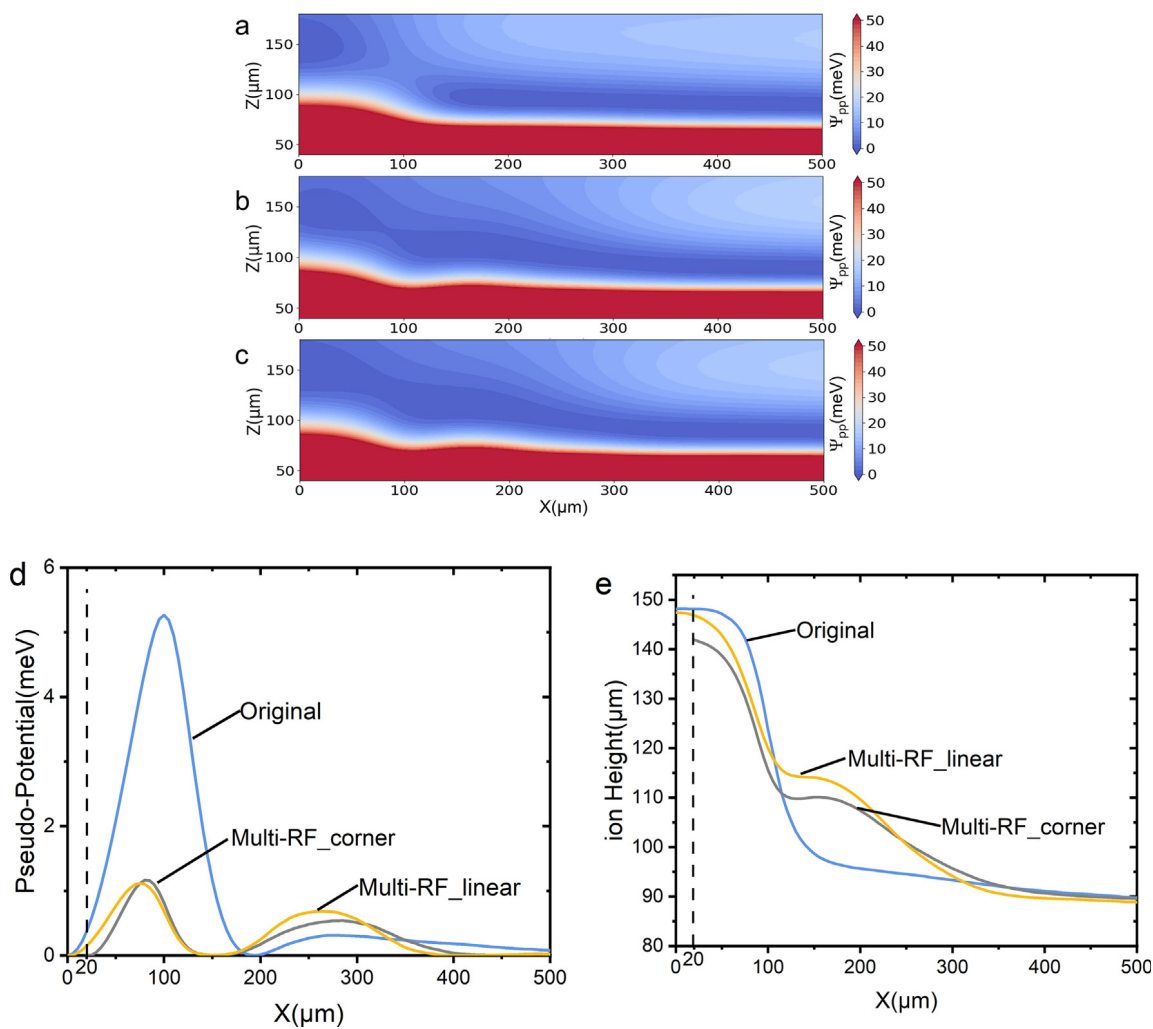

**Fig. 2 | The optimization results of the multi-RF field method. a–c**, The pseudo-potential distribution in the contour map for the ZOX ($y = 0$) plane; the $x$-axis stands for the distance from the junction center (origin) in zone A; the $z$-axis stands for the ion height from the surface electrode; the color stands for the value of the pseudo-potential; the brighter the color, the higher the potential value. **a**, Without optimization, all RF electrode voltages are set to 100 V. **b**, For the corner turning case, after the optimization of the RF voltages. **c**, For the linear shuttling case, after the optimization of the RF voltages. **d**, The relation between pseudo-potential and the $x$-axis position. The maximum pseudo-potential value for the original unoptimized case is is 5.265 meV (light blue), for corner turning it is 1.164meV (grey), and for linear shuttling it is 1.117 meV (orange). **e**, The relation between ion height and the $x$ position. 58.44 μm is the change in ion height for the original unoptimized case (light blue), 54.19 μm is for the corner turning optimization case (grey), and 58.59 um is for the linear shuttling case (orange). Noticing that there are no saddle point data in the range $x \in [0, 20]$ μm for corner turning mode because the pseudo-potential tube near $x \in [0, 20]$ μm starts to be off the $x$-axis and in a curve shape. RF, radio frequency.

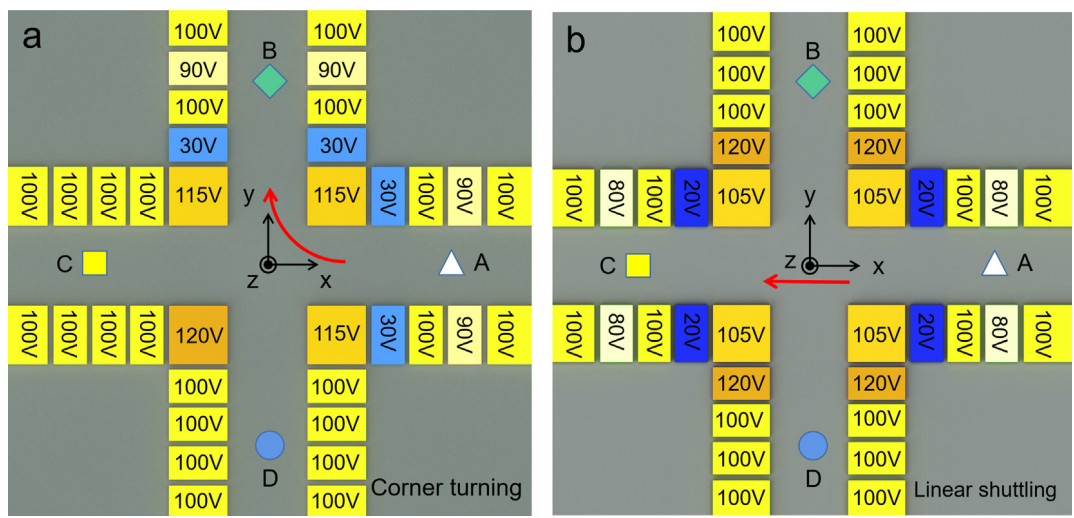

**Fig. 3 | The optimized voltage distribution of the multi-RF electrodes was achieved by using the multi-RF optimization method**. **a**, For the corner turning shuttling case, ions from zone A to zone B. **b**, For the linear shuttling case, ions from zone A to zone C.

In the case of linear shuttling, without loss of generality, it is assumed that the ions move from zone A to zone C. The result of the optimized pseudo-potential achieved by adopting the multi-RF field method is shown in Fig. 2c, and the corresponding RF voltage distribution is displayed in Fig. 3b. The RF voltages and the resulting trapping potential are symmetrical about $x = 0$. Compared to the original unoptimized result, the maximum pseudo-potential value has been greatly suppressed to 1.117 meV (orange solid line in Fig. 2d), and the maximum ion height variation has also been decreased to 58.59 μm (orange solid line in Fig. 2e).

From the above results, multiple RF fields with different amplitudes, the same frequency and phase can control and reduce the pseudo-potential barrier and the height variation of the ion near the junction without changing the shape of the electrodes. Both the corner turning and the linear shuttling cases achieve a much lower pseudo-potential barrier and a smaller ion height variation. This verified that the multi-RF field method does show the same optimizing effects and function as the geometrical shape optimization method. However, the RF barrier can be further optimized by shortening the distance between the junction center and the nearest electrode, reducing the separation between the RF electrode segments, or increasing the number of electrode segmentation, all of which will be beneficial for improving the field-controlling resolution and increasing the number of free parameters of the controlling RF fields[37,41].

On the other hand, in the process of the simulation, the distributed RF voltage optimization method only needs to solve the basis function once[48], while the traditional geometrical shape optimization method will solve the basis function in each optimizing iteration for a different shape of the electrodes. The finite element method (FEM) or boundary elemental method (BEM)[50] is required for precisely solving the basic function of the complex-shaped electrode, and setting large numbers of the geometrical free parameters tends to be the most time-consuming process for the optimization. Consequently, the multi-RF field optimization method can greatly shorten the time consumption. In our experience, rapid convergence of distributed RF electrode voltages can only be achieved by artificially searching voltage parameters within a few hours.

## THE HYBRID OPTIMIZATION

In the last section, it was verified that the potential barrier near the junction can be reduced by applying multiple RF fields with proper amplitudes, and a larger distance between the corner electrode and the junction center can yield weak field adjustability. Thus, in this section, the geometrical optimization and the multi-RF fields methods were hybridized by lengthening and reshaping the electrode corner shape, adding new RF electrode segments, then optimizing the RF voltage distribution. The final RF barrier can be further optimized and reduced.

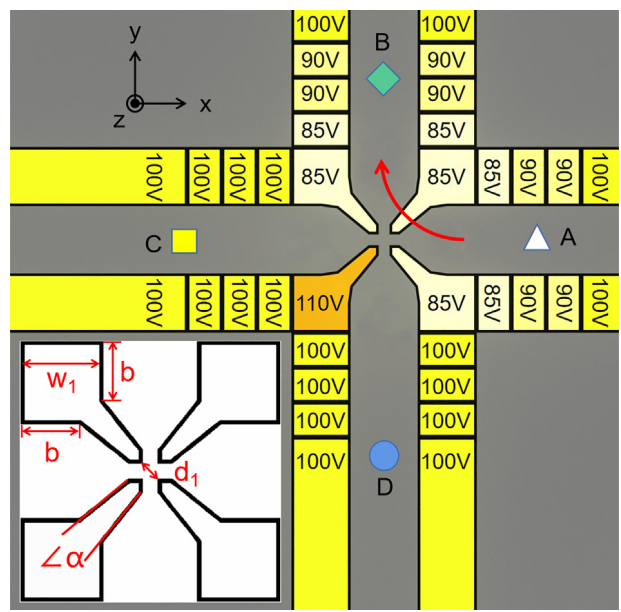

**Fig. 4 | The geometrical structure of the finger electrode and the final RF voltage distribution for the corner turning shuttling combine the optimized finger electrode and the optimized multi-RF voltages**. Keeping the same dimensions as in Fig. 1, the width of the electrode is $w_1$, the bottom position of the finger shape is $b$, the top angle of the finger is $\alpha$, which is symmetric about the $x = y$, and the distance between diagonal fingertips is $d_1$. In the numerical simulation process, to fix the dimensions $w_1 = 80$ μm and $b = 60$ μm, the final optimal electrode geometry parameters are $\alpha = 12.6°$ and $d_1 = 34$ μm. RF, radio frequency.

**Corner turning shuttling** To further reduce the maximum value of the pseudo-potential barrier in the corner turning shuttling case, the geometry optimization of the corner RF electrodes needs to be introduced. Change and optimization of the electrode geometry was firstly conducted, and subsequently followed by the optimization of the RF voltage distribution. The geometrical structure and the dimension of the new electrode are drawn in Fig. 4, extending the inner corner electrode length and forming the finger shape. Without loss of generality, the ions are assumed to transport from zone A to zone B. In the numerical simulation, the RF-driven frequency $\Omega_{RF}$ is 30 MHz, and all the RF voltages are fixed to 100 V when performing the geometrical optimization.

The optimized finger geometries for minimized pseudo-potential are as follows: the top angle $\alpha = 12.6°$, the bottom position $b = 60$ μm, and the distance between two diagonal fingers $d_1 = 34$ μm. The corresponding pseudo-potential and the ion height variation results are shown in Fig. 5. The minimum pseudo-potential is 0.757 meV, and the ion height variation is 7.37 μm (green solid line). By applying the multi-RF field optimization method to the new structure of the electrodes, the RF barrier can be further reduced to 0.136 meV, and the ion height variation is 14.65 μm (orange solid line). The final RF voltage distribution is shown in Fig. 4, symmetric about $x = y$. The intersection of the symmetry plane $x = y$ and the center of the curved pseudo-potential tube is around $(x, y) = (20, 20)$ μm. Near the junction center, the saddle

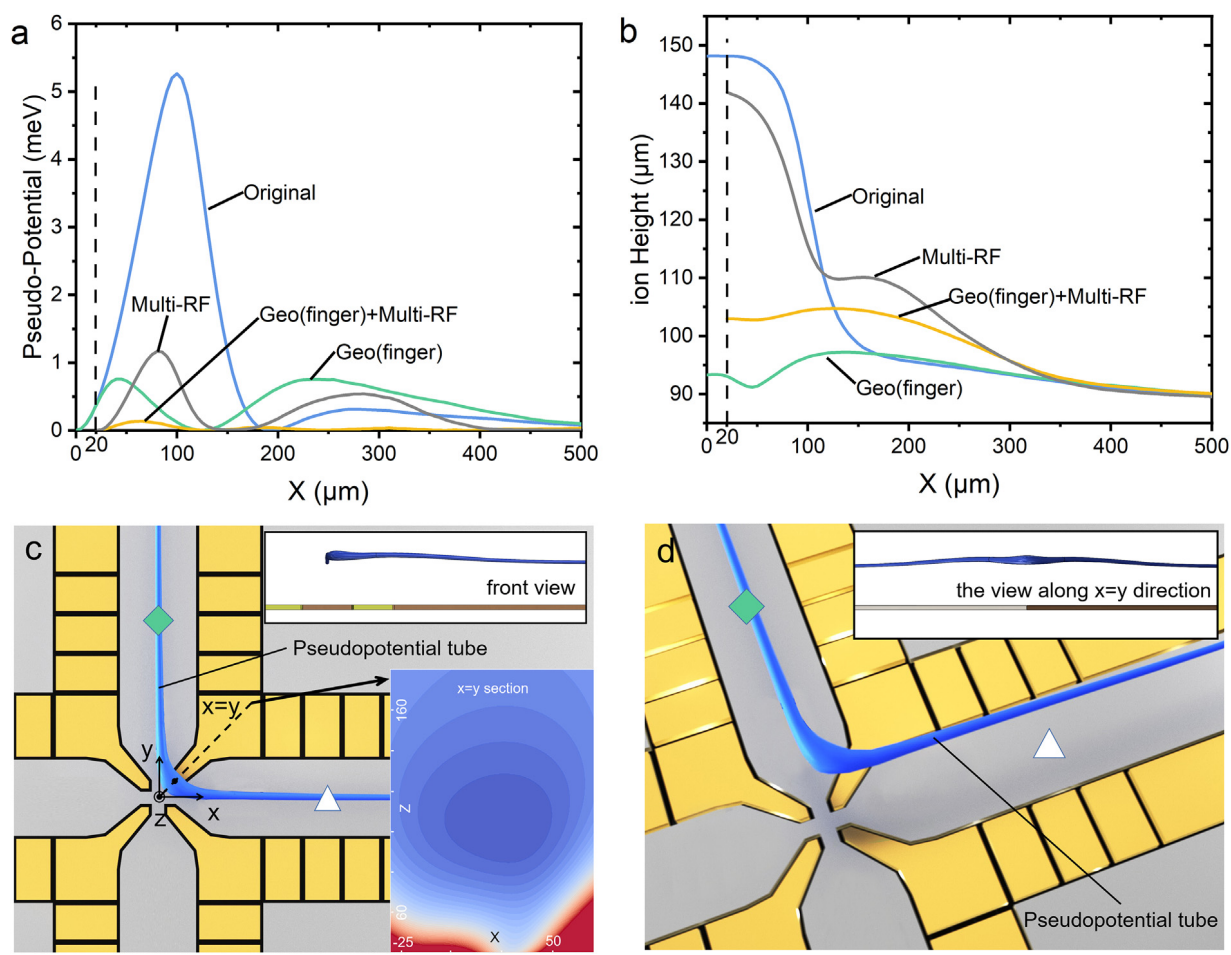

**Fig. 5 | Optimization results for corner turning shuttling. a**, The pseudo-potential distribution along the *x*-axis in zone A. **b**, The ion height distribution along the *x*-axis in zone A. The results for the original geometrical shape with all RF voltages 100V (light blue), the original geometrical shape with optimized RF voltage distribution (grey), the optimized finger shape electrode with all RF voltages 100 V (green), and the optimized finger shape electrode with optimized RF voltage (orange). The final RF voltage distribution (orange) is shown in Fig. 4. The potential barrier and the ion height variation for different optimized situations are 5.265 meV and 58.44 μm (light blue), 1.164 meV and 54.19 μm (grey), 0.757 meV and 7.37 μm (green), and 0.136 meV and 14.65 μm (orange). There is no saddle point data in the range of $x \in [0, 20]$ μm for corner turning mode because the pseudo-potential tube near $x \in [0, 20]$ μm starts to be off the *x*-axis and in a curve shape (grey and orange). **c, d**, Different views of the 3D schematic diagram of the pseudo-potential tube obtained with $\Psi = 0.400$ meV as the outer wall. **c**, The top view and front view; the inset is the pseudo-potential distribution along the symmetric plane $x = y$. **d**, The oblique view, and the view along the $x = y$ line. The tube diameter is larger near the center of the junction and smaller towards the linear trap region. RF, radio frequency.

points are off the *x*-axis, thus there are no saddle point data in the range of x∈[0, 20] μm.

The above results show that the minimum RF barriers obtained by the multi-RF fields method and the geometrical optimization method are similar. The effects of the two methods are consistent and can be replaced by each other. The combination of these two methods simplifies the complexity of the electrode geometry while maintaining the minimized value of the pseudo-potential. In order to make an intuitive comparison, the optimized pseudo-potential tube for corner-turning shuttling is shown from different perspectives in Fig. 5c and d. The pseudo-potential tube is an isosurface with a pseudo-potential value of 0.4 meV, continuous, flat, and symmetrical about the $x = y$. Near the center of the junction, the pseudo-potential is still a tube shape, and the trapping frequency and the trap depth of the junction center are at the same level as the linear regions. The diameter of the pseudo-potential tube is about 3 times that of the pseudo-potential tube in the linear trapping region. In the inset of Fig. 5c, the pseudo-potential distribution on the $x = y$ section is shown.

**Linear shuttling**   Based on the optimized finger shape of the electrodes and the RF voltage distribution results in the previous section, the intuitive way to generate a better trapping potential is to simply apply the optimized RF voltages in zone B to zone C and apply the 100 V RF voltage to the electrode segments in zone B. But this intuitive method can't greatly improve the trapping potential. The corresponding pseudo-potential (green line in Fig. 7a) is similar to the potential without the finger shape optimization (orange line in Fig. 2d). As a comparison, the results of pseudo-potential and ion height variation of the original electrode pattern after applying 100 V RF voltage and multi-RF field optimization are repeatedly plotted in Fig. 7, respectively, as shown by the light blue and orange curves. The potential barrier of the intuitive method is 0.681 meV, and the ion height variation is 18.47 μm.

Similar to the linear trap electrodes, supposing that ion is still transported from region A to region C, two pairs of wedge electrodes (RF1f and RF2f in Fig. 6) can be introduced in the middle of the finger electrode to compensate for the interrupted linear ion trap electrode. The shape of the introduced wedge electrode is shown in Fig. 6, with width $w_2$, length $l_2$, apex angle $\beta$, and spacing $d_2$. During the numerical simulation, the apex angle $\beta = 53°$ was fixed, and the wedge electrode geometry parameters $w_2$, $l_2$ and $d_2$ were optimized. To ensure the continuity of the electric field, the applied RF voltage on the RF2f electrode was set to 85 V during optimization. The final results after optimization are $w_2 = 29$ μm, $l_2 = 40$ μm, and $d_2 = 152$ μm, and the corresponding pseudo-potential distribution and the ion height distribution are shown in Fig. 7a and b with the grey line. After introducing the wedge electrode, the pseudo-potential barrier was obviously reduced to 0.165 meV, and the ion height variation was also reduced to 14.93 μm. Significant improvement hasn't been achieved by further optimization of the RF distribution voltage with the corresponding pseudo-potential being as 0.128 meV, and the ion height variation as 14.97 μm. The pseudo-potential distribution on the $x = 0$ section is shown in the inset of Fig. 7c.

## DISCUSSION

The above results verified that the multi-RF field method can both reduce the potential barrier and flatten the pseudo-potential center height near the junction. However, the optimized RF voltages for corner turning and linear shuttling are different. Thus, the shuttling of the ions in two directions cannot be implemented with the same set of voltages. A switching protocol is essential.

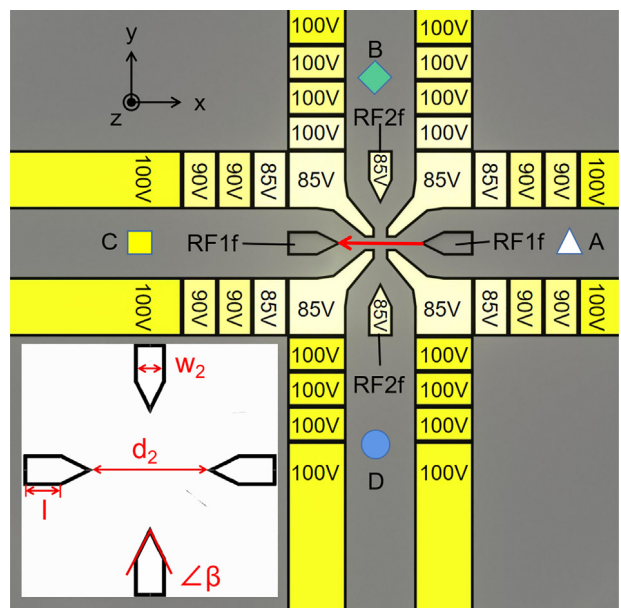

**Fig. 6 | The electrode pattern, optimized voltage distribution, and wedge electrode geometric parameters changed after the introduction of wedge electrodes.** The finger electrodes were kept the same size as in Fig. 4. The geometrical parameters of the wedge electrodes are the top angle $\beta$, the width $w_2$, the length $l$, and the distance between two wedge diagonal apexes. For the numerical simulation, we fixed the top angle $\beta = 53°$, and the optimized electrode geometry parameters are: $w_2 = 29$ μm, $l = 40$ μm, $d_2 = 152$ μm.

Several studies have been conducted involving the use of multiple RF electric fields to control the trapping pseudo-potential[44] and auxiliary ion shuttling. But the saddle point must be moved along with the ions during the ion shuttling process. It is a complex controlling process, not only to design a proper saddle point at a certain position, but the saddle point position is time-dependent, and the RF fields and the DC shuttling fields must be matched precisely. In addition, the variable of the RF voltage over a short time may contribute to the temperature fluctuation of the electrodes, changing the capacitance between electrodes and thus causing the resonance frequency drift, which increases the difficulty and instability of the experiment.

Different from previous researches, the multi-RF field dynamic RF voltage controlling method proposed in the current work is quasi-static. The voltage switching process is determined by the shuttling program with the quantum algorithm. It exhibits a much slower voltage-changing frequency than dynamic ways and separates the RF trapping-potential controlling process and the DC ion shuttling-controlling process. This simplifies the electric field-controlling waveform, reduces the RF power and electrode temperature fluctuations, and improves the stability of the system.

From an experimental point of view, our method is also feasible. Although multiple channels of RF sources are required, there are only four independent RF voltage channels demands for each junction. Any possible shuttling direction is taken into consideration, such as corner turning or linear shuttling, as shown in Fig. 8. Since the RF voltage switching frequency is slow, the RF source number can be further reduced when combined with the latch. Thus, our method can be applied to the ion trap chip design for scalable quantum computing experiments.

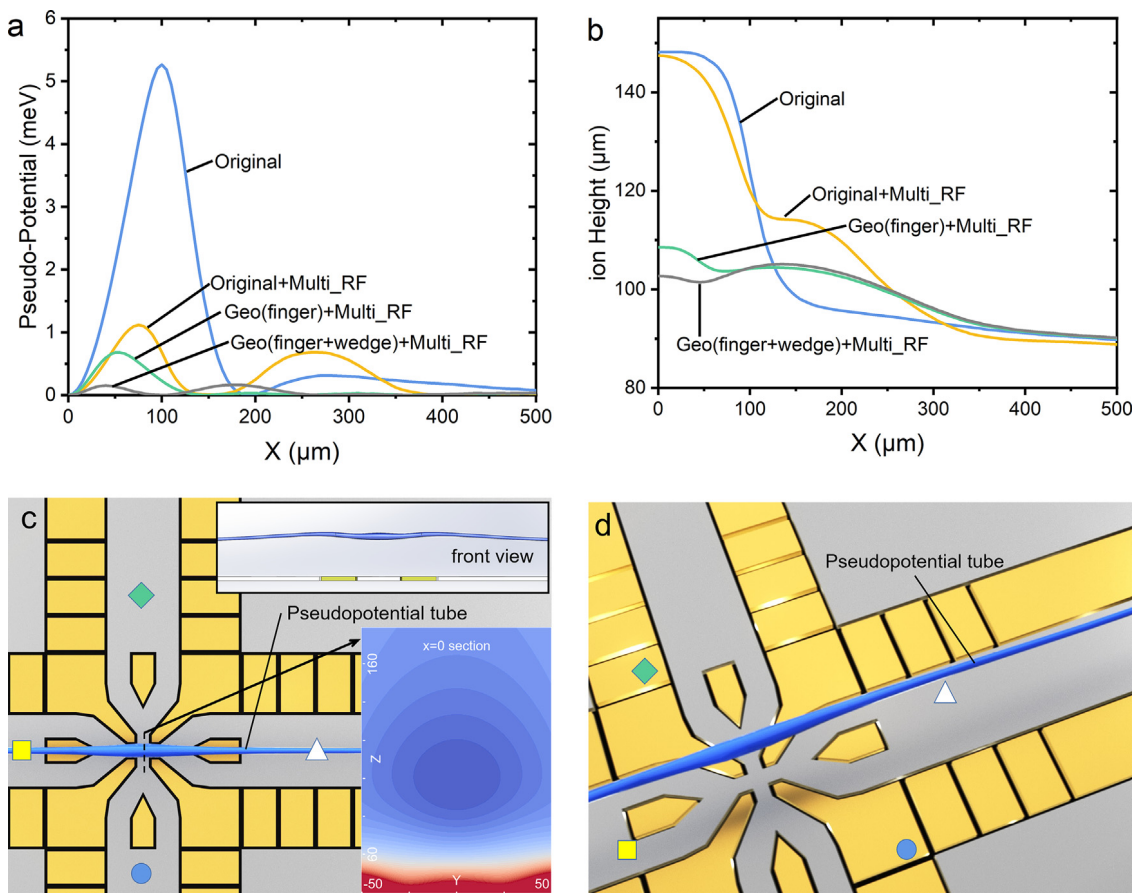

**Fig. 7 | Optimization results for linear shuttling**. **a**, The pseudo-potential distribution along the *x*-axis in zone A. **b**, The ion height distribution along the *x*-axis in zone A. The results for the original geometrical shape with all RF voltages 100V (light blue, the same as the light blue in Fig. 2d and e), the original geometrical shape with optimized RF voltage distribution for linear shuttling (orange, the same as the orange line in Fig. 2d and e), the optimized finger shape electrode with optimized RF voltage for corner turning shuttling (green), and the optimized finger and wedge shape electrodes with optimized RF voltage (grey). The final RF voltage distribution (grey) is shown in Fig. 6. The potential barrier and the ion height variation for different optimized situations are 5.265 meV and 58.44 μm (light blue), 1.117 meV and 58.59 μm (orange), 0.681 meV and 18.47 μm (green), and 0.165 meV and 14.93 μm (grey). **c, d**, Different views of the 3D schematic diagram of the pseudo-potential tube obtained with Ψ = 0.4 meV as the outer wall. **c**, The top view and front view; the inset is the pseudo-potential distribution on the plane *x* = 0. **d**, The oblique view. The tube diameter is larger near the center of the junction and smaller toward the linear trap region. RF, radio frequency.

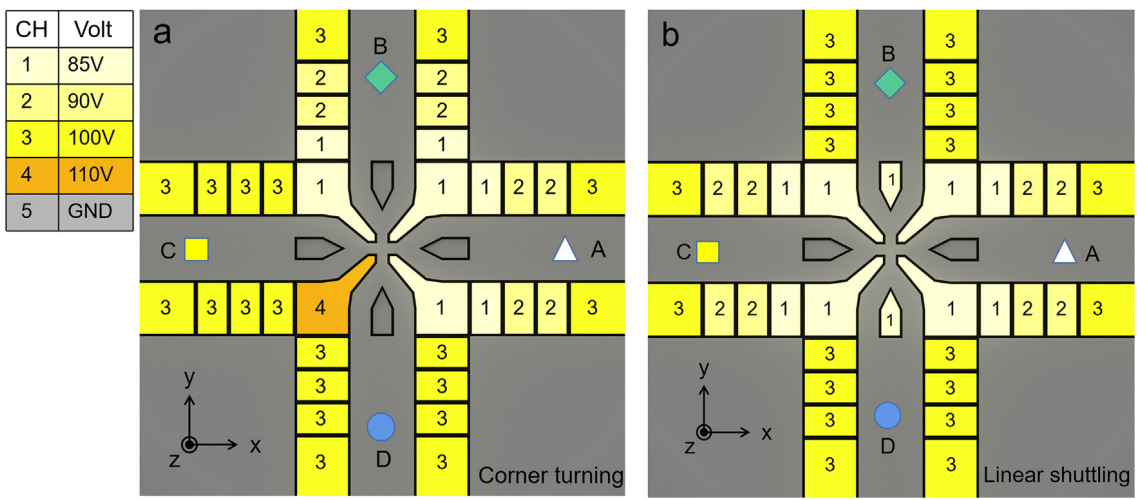

**Fig. 8 | RF voltage loading scheme corresponding to two shuttling modes**. **a**, RF voltage loading scheme for the corner turning shuttling mode. **b**, RF voltage loading scheme for the linear shuttling mode. The number on each RF segment presents an individual RF channel, and the table on the left gives the voltage values for each different RF channel. Noticing that, in each mode, there are at most four independent RF channels. RF, radio frequency.

On the other hand, the optimized pseudo-potential tube in the current work is endowed with the advantage of potentially increasing the shuttling speed for corner-turning cases. In the traditional method, the turn shuttle must be firstly slowed down, stopped in the center of the junction, and then accelerated in the second direction, changing the direction of motion from X to Y. However, the optimized corner-turning shuttling pseudo-potential in this study is in a curved shape, and it can potentially accelerate the corner-shuttling speed along a curved trajectory without decreasing or losing the linear velocity.

## CONCLUSION

In summary, this study presented a method to modulate the pseudo-potential near the junction of an ion trap chip with the adoption of multiple RF fields. An RF electrode was divided into multiple sub-electrodes, and the customized optimization of the confinement electric field was realized by applying the RF voltage with the same frequency, the same phase, and a different amplitude to each sub-electrode. The effectiveness of the method and its equivalence with electrode pattern optimization were illustrated. Better pseudo-potential distribution and ion height variation results can be obtained when combining multi-RF voltage distribution optimization with graphical optimization, and the feasibility of the experiment was also discussed.

The all-electrical method of updating the RF voltage amplitude to control the confinement electric field can simplify the design of the electrode pattern, reduce the difficulty of processing, and reduce the chip discharge breakdown caused by the sharp electrode. This study also provides a method to adjust the distribution of the imprisoned potential field even after determining the electrode shape, which provides the possibility for investigating two-dimensional ion crystals and their phase transitions.

## MISCELLANEA

**Acknowledgements** The authors want to thank supported from the Natural Science Foundation of Guizhou Province of China (Grants No. ZCKJ2021015), Guizhou Science Platform and Talent (Grants No. GCC[2023]090). This work was supported by Guangdong Basic and Applied Basic Research Foundation (Grants No. 2020A1515010864) and the National Natural Science Foundation of China (Grants No. 11904423).

**Declaration of competing interest** The authors declare no competing interests.